\documentclass{llncs}
\usepackage{url}
\usepackage{amsmath}
\usepackage[all]{xy}

\newcommand{\Pro}{\ensuremath{\mathbf{Pr}}}

\usepackage{hyperref}
\hypersetup{ 
  bookmarksnumbered, 
  pdftitle={Learning in Multiagent Systems: An Introduction from a Game-Theoretic Perspective},
  pdfauthor={Jose M. Vidal},
  pdfsubject={},
  pdfkeywords={},
  pdfview=FitH,
  pdfstartview=FitH 
}

\usepackage{fancyhdr}

\begin{document}
\mainmatter              

\title{Learning in Multiagent Systems: An Introduction from a Game-Theoretic Perspective}

\titlerunning{Learning in Multiagent Systems: An Introduction}

\author{Jos\'{e} M. Vidal}
\tocauthor{Jos\'{e} M. Vidal (University of South Carolina)}
\institute{ 
University of South Carolina, Computer Science and Engineering,\\
Columbia, SC 29208 \\
\email{vidal@sc.edu} }
\authorrunning{J. Vidal}   
\maketitle

\begin{abstract}
  We introduce the topic of learning in multiagent systems. We first
  provide a quick introduction to the field of game theory, focusing
  on the equilibrium concepts of iterated dominance, and Nash
  equilibrium.  We show some of the most relevant findings in the
  theory of learning in games, including theorems on fictitious play,
  replicator dynamics, and evolutionary stable strategies. The CLRI
  theory and n-level learning agents are introduced as attempts to
  apply some of these findings to the problem of engineering
  multiagent systems with learning agents. Finally, we summarize some
  of the remaining challenges in the field of learning in multiagent
  systems.
\end{abstract}

\thispagestyle{fancy}
\lhead{}
\chead{In Eduardo Alonso, editor, Adaptive Agents: LNAI 2636. Springer Verlag, 2003.}
\cfoot{\copyright{} Springer Verlag, 2003.}

\section{Introduction}
\label{sec:introduction}

The engineering of multiagent systems composed of learning agents
brings together techniques from machine learning, game theory, utility
theory, and complex systems. A designer must choose carefully which
machine-learning algorithm to use since otherwise the system's
behavior will be unpredictable and often undesirable. Fortunately, we
can use the tools from these areas in an effort to predict the
expected system behaviors. In this article we introduce these
techniques and explain how they are used in the engineering of
learning multiagent systems.

The goal of machine learning research is the development of algorithms
that increase the ability of an agent to match a set of inputs to
their corresponding outputs \cite{mitchell97a}. That is, we assume the
existence of a large, sometimes infinite, set of examples $E$. Each
example $e \in E$ is a pair $e = \{a, b\}$ where $a \in A$ represents
the input the agent receives and $b \in B$ is the output the agent
should produce when receiving this input. The agent must find a
function $f$ which maps $A \rightarrow B$ for as many examples of $A$
as possible.  In a controlled test the set $E$ is usually first
divided into a \emph{training set} which is used for training the
agent, and a \emph{testing set} which is used for testing the
performance of the agent. In some scenarios it is impossible to first
train the agent and then test it. In these cases the training and
testing examples are interleaved. The agent's performance is assessed
on an ongoing manner.

When a learning agent is placed in a multiagent scenario these
fundamental assumptions of machine learning are violated. The agent is
no longer learning to extrapolate from the examples it has seen of
fixed set $E$, instead it's target concept keeps changing, leading to
a moving target function problem \cite{vidal:98a}. In general,
however, the target concept does not change randomly; it changes based
on the learning dynamics of the other agents in the system. Since
these agents also learn using machine learning algorithms we are left
with some hope that we might someday be able to understand the complex
dynamics of these type of systems.

Learning agents are most often selfish utility maximizers. These
agents often face each other in encounters where the simultaneous
actions of a set of agents leads to different utility payoffs for all
the participants. For example, in a market-based setting a set of
agents might submit their bids to a first-price sealed-bid auction.
The outcome of this auction will result in a utility gain or loss for
all the agents. In a robotic setting two agents headed in a collision
course towards each other have to decide whether to stay the course or
to swerve. The results of their combined actions have direct results
in the utilities the agents receive from their actions. We are solely
concerned with learning agents that maximize their own utility. We
believe that systems where agents share partial results or otherwise
help each other can be considered extension on traditional machine
learning research.

\section{Game Theory}
\label{sec:game-theory}

Game theory provides us with the mathematical tools to understand the
possible strategies that utility-maximizing agents might use when
making a choice. It is mostly concerned with modeling the decision
process of rational humans, a fact that should be kept in mind as we
consider its applicability to multiagent systems.

The simplest type of game considered in game theory is the
\emph{single-shot simultaneous-move} game. In this game all agents
must take one action. All actions are effectively simultaneous. Each
agent receives a utility that is a function of the combined set of
actions. In an \emph{extended-form} game the players take turns and
receive a payoff at the end of a series of actions. A single-shot game
is a good model for the types of situations often faced by agents in a
multiagent system where the encounters mostly require coordination.
The extended-form games are best suited to modeling more complex
scenarios where each successive move places the agents in a different
state. Many scenarios that first appear like they would need an
extended-form game can actually be described by a series of
single-shot games. In fact, that is the approach taken by many
multiagent systems researchers.

In the one-shot simultaneous-move game we say that each agent $i$
chooses a \emph{strategy} $s_i \in S_i$, where $S_i$ is the set of all
strategies for agent $i$. These strategies represent the actions the
agent can take. When we say that $i$ chooses strategy $s_i$ we mean
that it chooses to take action $s_i$. The set of all strategies chosen
by all the agents is the \emph{strategy profile} for that game and it
is denoted by $s \in S \equiv \times_{i=i}^I S_i$. Once all the agents
make their choices and form the strategy profile $s$ then each agent
$i$ receives a \emph{utility} which is given by the function $u_i(s)$.
Notice that a player's utility depends on the choices made by all the
agents.

Two player games involve only two players, $i$ and $j$. They are often
represented using a game matrix such as the one shown in
Figure~\ref{fig:gamematrix}. In that matrix we see that if agent 1
(the one who chooses from the rows) chooses action A and agent 2
chooses action B then agent 1 will receive a utility of 3 while agent
2 receives a utility of 4. Using our notation for strategies we would
say that if the strategy profile is $(s_1, s_2)$ then the payoff
vector is
\[
(u_1(s_1,s_2), u_2(s_1,s_2))
\]

\begin{figure}[tb]
  \begin{center}
    \begin{tabular}{r@{}c@{}c@{}}
      & A & B \\ \cline{2-3}
      A\ \  &\vline\ 1,2\ \ &\vline\ 3,4 \vline \\ \cline{2-3}
      B\ \  &\vline\ 3,2\ \ &\vline\ 2,1 \vline \\ \cline{2-3}
    \end{tabular}
  \end{center}
  \caption{Sample two-player game matrix. Agent 1 chooses from the
    rows and agent 2 chooses from the columns.} 
  \label{fig:gamematrix}
\end{figure}

It is possible that a player will choose randomly between its action
choices, using different prior probabilities for each choice. These
types of strategies are called \emph{mixed strategies} and they are a
probability distribution over an agent's actions. We say that a mixed
strategy for agent $i$ is $\sigma_i \in \Sigma_i \equiv P(S_i)$ where
$P(S_i)$ is the set of all probability distributions over the set of
pure strategies $S_i$. Although a real agent can not take a ``mixed
action'', mixed strategies are useful abstractions since they allow us
to model agents who might use some randomization subroutine to choose
their action.

\section{Solution Concepts}
\label{sec:solution-concepts}

Much of the work in game theory has concentrated in the definition of
plausible solution concepts. A solution concept tries to define the
set of actions that a set of rational agents will choose when faced
with a game. The most common assumptions are that the agents are
rational, have common knowledge\footnote{Common knowledge about $p$
  means that everybody knows that everybody knows, and so on to
  infinity, about $p$.} of the payoffs in the game matrix, and that
they are intelligent enough to re-create the thought process that the
mathematician went through to come up with the solution concept. As
such, most solution concepts are geared towards an understanding of
how smart, well-informed people would act. They are not necessarily
meant to explain the behavior of machine-learning agents. Still, the
fact that they provide the ``best'' solution makes them a useful tool.

\subsection{Iterated Dominance}
\label{sec:iterated-dominance}

The iterated dominance approach is to successively eliminate from
consideration those actions that are worst than some other action, no
matter what the other player does. For example, in Figure~\ref{fig:id}
we see a game where agent 1's action B is dominate by A. That is, no
matter what agent 2 does, agent 1 should choose action A. Then, if
agent 1 chooses action A then agent 2 should choose action B.
Therefore, the solution strategy profile for this game is $(A,B)$.

\begin{figure}[tb]
  \begin{center}
    \begin{tabular}{r@{}c@{}c@{}}
      & A & B \\ \cline{2-3}
      A\ \  &\vline\ 8,2\ \ &\vline\ 9,4 \vline \\ \cline{2-3}
      B\ \  &\vline\ 1,2\ \ &\vline\ 3,1 \vline \\ \cline{2-3}
    \end{tabular}
  \end{center}
  \caption{A game where agent 1's action B is dominated by A.}
  \label{fig:id}
\end{figure}

Formally, we say that a strategy $\sigma_i$ is \emph{strictly
  dominated} for agent $i$ if there is some other strategy
$\tilde{\sigma}_i \in \Sigma_i$ for which $u_i(\tilde{\sigma}_i,
\sigma_{-i}) > u_i(\sigma_i, \sigma_{-i})$ for all $\sigma_{-i}$,
where $\sigma_{-i}$ is a set of strategies for all agents except
$i$. Notice that the inequality sign is a greater-than. If we change
that sign to a greater-than-or-equal then we have the definition for a
\emph{weakly dominated} strategy. 

There is no reason for a rational agent to choose a strictly dominated
strategy. That is, there is no reason for an agent to choose
$\sigma_i$ when there exists a $\tilde{\sigma}_i$ which will give it a
better utility no matter what the other agents do. Similarly, there is
no reason for the agent to choose a weakly dominated strategy. Of
course, this reasoning relies on the assumption that the agent can
indeed determine the existence of a $\tilde{\sigma}_i$. This
assumption can be hard to justify in cases where the better strategy
is a mixed strategy where the agent has an infinite number of possible
strategies to verify, or in cases where the number of actions and
agents is too large to handle.

The iterated dominance algorithm consists of calculating all the
strategies that are dominated for all the players, eliminating those
strategies from consideration, and repeating the process until no more
strategies are dominated. At that point it might be the case that only
one strategy profile is left available. In this case that profile is
the one all agents should play. However, in many cases the algorithm
still leaves us with a sizable game matrix with a large number of
possible strategy profiles. The algorithm then serves only to reduce
the size of the problem.

\subsection{Nash Equilibrium}
\label{sec:nash-equilibrium}

The Nash equilibrium solution concept is popular because it provides a
solution where other solution concepts fail. The Nash equilibrium
strategy profile is defined as $\hat{\sigma}$ such that for all agents
$i$ it is true that there is no strategy better than $\hat{\sigma}_i$
given that all the other agents take the actions prescribed by
$\hat{\sigma}_{-i}$.  Formally, we say that $\hat{\sigma}$ is a Nash
equilibrium strategy profile if for all $i$ it is true that
$\hat{\sigma}_i \in BR_i(\hat{\sigma_{-i}})$, where $BR_i(s_{-i})$ is
the best response for $i$ to $s_{-i}$.  That is, given that everyone
else plays the strategy given by the Nash equilibrium the best
strategy for any agent is the one given by the Nash equilibrium.  A
strict Nash equilibrium states that $\hat{\sigma}_i$ is strictly
(i.e., greater than) better than any other alternative.

It has been shown that every game has at least one Nash equilibrium,
as long as mixed strategies are allowed. The Nash equilibrium has the
advantage of being stable under single agent desertions. That is, if
the system is in a Nash equilibrium then no agent, working by itself,
will be tempted to take a different action.  However, it is possible
for two or more agents to conspire together and find a set of actions
which are better for them. This means that the Nash equilibrium is not
stable if we allow the formation of coalitions.

Another problem we face when using the Nash equilibrium is the fact
that a game can have multiple Nash equilibria. In these cases we do
not know which one will be chosen, if any. The Nash equilibrium could
also be a mixed strategy for some agent while in the real world the
agent has only discrete actions available. In both of these cases the
Nash equilibrium is not sufficient to identify a unique strategy
profile that rational agents are expected to play. As such, further
studies of the dynamics of the system must be carried out in order to
refine the Nash equilibrium solution. The theory of learning in
games---a branch of game theory---has studied how simple learning
mechanisms lead to equilibrium strategies.

\section{Learning in Games}
\label{sec:learning-games}

The theory of learning in games studies the equilibrium concepts
dictated by various simple learning mechanisms. That is, while the
Nash equilibrium is based on the assumption of perfectly rational
players, in learning in games the assumption is that the agents use
some kind of algorithm. The theory determines the equilibrium strategy
that will be arrived at by the various learning mechanisms and maps
these equilibria to the standard solution concepts, if
possible. Many learning mechanisms have been studied. The most common
of them are explained in the next few sub-sections.

\subsection{Fictitious Play}
\label{sec:fictitious-play}

A widely studied model of learning in games is the process of
fictitious play. In it agents assume that their opponents are playing
a fixed strategy. The agents use their past experiences to build a
model of the opponent's strategy and use this model to choose their
own action. Mathematicians have studied these types of games in order
to determine when and whether the system converges to a stable
strategy.

Fictitious play uses a simple form of learning where an agent
remembers everything the other agents have done and uses this
information to build a probability distribution for the other agents'
expected strategy. Formally, for the two agent ($i$ and $j$) case we
say that $i$ maintains a weight function $k_i: S_{j} \rightarrow
\mathcal{R}^+$. The weight function changes over time as the agent
learns. The weight function at time $t$ is represented by $k_i^t$
which keeps a count of how many times each strategy has been played.
When at time $t-1$ opponent $j$ plays strategy $s_j^{t-1}$ then $i$
updates its weight function with
\begin{equation}
  \label{eq:1}
k_i^t(s_j) = k_i^{t-1}(s_j) + 
\left\{
    \begin{array}{ll}
      1 & \mbox{if $s_j^{t-1} = s_j$,} \\
      0 & \mbox{if $s_j^{t-1} \neq s_j$.}
    \end{array}    
\right.
\end{equation}

Using this weight function, agent $i$ can now assign a probability to
$j$ playing any of its $s_j \in S_j$ strategies with
\begin{equation}
  \label{eq:2}
\Pro_i^t[s_j] = \frac{k_i^t(s_j)}{\sum_{\tilde{s}_j \in S_j} k_i^t(\tilde{s}_j)}.
\end{equation}

Player $i$ then determines the strategy that will give it the highest
expected utility given that $j$ will play each of its $s_j \in S_j$
with probability $\Pro_i^t[s_j]$. That is, $i$ determines its best
response to a probability distribution over $j$'s possible strategies.
This amounts to $i$ assuming that $j$'s strategy at each time is taken
from some fixed but unknown probability distribution.

Several interesting results have been derived by researches in this
area.  These results assume that all players are using fictitious
play. In \cite{fudenberg90a} it was shown that the following two
propositions hold.

\begin{proposition}
  If $s$ is a strict Nash equilibrium and it is played at time $t$
  then it will be played at all times greater than $t$.
\end{proposition}

Intuitively we can see that if the fictitious play algorithm leads to
all players to play the same Nash equilibrium then, afterward, they
will increase the probability that all others are playing the
equilibrium. Since, by definition, the best response of a player when
everyone else is playing a strict Nash equilibrium is to play the same
equilibrium, all players will play the same strategy and the next
time. The same holds true for every time after that.

\begin{proposition}
  If fictitious play converges to a pure strategy then that strategy
  must be a Nash equilibrium.
\end{proposition}

We can show this by contradiction. If fictitious play converges to a
strategy that is not a Nash equilibrium then this means that the best
response for at least one of the players is not the same as the
convergent strategy. Therefore, that player will take that action at
the next time, taking the system away from the strategy profile it was
supposed to have converged to.

\begin{figure}[tb]
  \begin{center}
    \begin{tabular}{r@{}c@{}c@{}}
      & A & B \\ \cline{2-3}
      A\ \  &\vline\ 0,0\ \ &\vline\ 1,1 \vline \\ \cline{2-3}
      B\ \  &\vline\ 1,1\ \ &\vline\ 0,0 \vline \\ \cline{2-3}
    \end{tabular}
  \end{center}
  \caption{A game matrix with an infinite cycle.}
  \label{fig:cycle}
\end{figure}

An obvious problem with the solutions provided by fictitious play can
be seen in the existence of infinite cycles of behaviors. An example
is illustrated by the game matrix in Figure~\ref{fig:cycle}. If the
players start with initial weights of $k_1^0(A)=1$, $k_1^0(B)= 1.5$,
$k_2^0(A)=1$, and $k_2^0(B)= 1.5$ they will both believe that the
other will play $B$ and will, therefore, play $A$. The weights will
then be updated to $k_1^1(A)=2$, $k_1^1(B)= 1.5$, $k_2^1(A)=2$, and
$k_2^1(B)= 1.5$. Next time, both agents will believe that the other
will play $A$ so both will play $B$. The agents will engage in an
endless cycle where they alternatively play $(A,A)$ and $(B,B)$. The
agents end up receiving the worst possible payoff.

This example illustrates the type of problems we encounter when adding
learning to multiagent systems. While we would hope that the machine
learning algorithm we use will be able to discern this simple pattern
and exploit it, most learning algorithms can easily fall into cycles
that are not much complicated than this one.  One common strategy for
avoiding this problem is the use of randomness. Agents will sometimes
take a random action in an effort to exit possible loops and to
explore the search space. It is interesting to note that, as in the
example from Figure~\ref{fig:cycle}, it is often the case that the
loops the agents fall in often reflect one of the mixed strategy Nash
equilibria for the game. That is, $(.5,.5)$ is a Nash equilibrium
for this game. Unfortunately, if the agents are synchronized, as in
this case, the implementation of a mixed strategy could lead to a
lower payoff.

Games with more than two players require that we decide whether the
agent should learn individual models of each of the other agents
independently or a joint probability distribution over their combined
strategies.  Individual models assume that each agent operates
independently while the joint distributions capture the possibility
that the others agents' strategies are correlated. Unfortunately, for
any interesting system the set of all possible strategy profiles is
too large to explore---it grows exponentially with the number of
agents. Therefore, most learning systems assume that all agents
operate independently so they need to maintain only one model per
agent.

\subsection{Replicator Dynamics}
\label{sec:replicator-dynamics}

Another widely studied model is \emph{replicator dynamics}.  This
model assumes that the percentage of agents playing a particular
strategy will grow in proportion to how well that strategy performs in
the population. A homogeneous population of agents is assumed. The
agents are randomly paired in order to play a symmetric game, that is,
a game where both agents have the same set of possible strategies and
receive the same payoffs for the same actions. The replicator dynamics
model is meant to capture situations where agents reproduce in
proportion to how well they are doing.

Formally, we let $\phi^t(s)$ be the number of agents using strategy
$s$ at time $t$. We can then define
\begin{equation}
  \label{eq:3}
\theta^t(s) = \frac{\phi^t(s)}{\sum_{s'\in S}\phi^t(s')}  
\end{equation}

to be the fraction of agents playing $s$ at time $t$. The expected
utility for an agent playing strategy $s$ at time $t$ is defined as 
\begin{equation}
  \label{eq:4}
u^t(s) \equiv \sum_{s' \in S} \theta^t(s')u(s,s'),
\end{equation}

where $u(s,s')$ is the utility than an agent playing $s$ receives
against an agent playing $s'$. Notice that this expected utility
assumes that the agents face each other in pairs and choose their
opponents randomly. In the replicator dynamics the reproduction rate
for each agent is proportional to how well it did on the previous
step, that is,
\begin{equation}
  \label{eq:5}
\phi^{t+1}(s) = \phi^t(s)(1 + u^t(s)).
\end{equation}

Notice that the number of agents playing a particular strategy will
continue to increase as long as the expected utility for that strategy
is greater than zero. Only strategies whose expected utility is
negative will decrease in population. It is also true that under these
dynamics the size of a population will constantly fluctuate. However,
when studying replicator dynamics we ignore the absolute size of the
population and focus on the fraction of the population playing a
particular strategy, i.e., $\theta^t(s)$, as time goes on.  We are
also interested in determining if the system's dynamics will converge
to some strategy and, if so, which one.

In order to study these systems using the standard solution concepts
we view the fraction of agents playing each strategy as a mixed
strategy for the game. Since the game is symmetric we can use that
strategy as the strategy for both players, so it becomes a strategy
profile. We say that the system is in a Nash equilibrium if the
fraction of players playing each strategy is the same as the
probability that the strategy will be played on a Nash equilibrium.
In the case of a pure strategy Nash equilibrium this means that all
players are playing the same strategy.

An examination of these systems quickly leads to the conclusion that
every Nash equilibrium is a steady state for the replicator dynamics.
In the Nash equilibrium all the strategies have the same average
payoff since the fraction of other players playing each strategy
matches the Nash equilibrium. This fact can be easily proven by
contradiction. If an agent had a pure strategy that would return a
higher utility than any other strategy then this strategy would be a
best response to the Nash equilibrium. If this strategy was different
from the Nash equilibrium then we would have a best response to the
equilibrium which is not the equilibrium, so the system could not be
at a Nash equilibrium.

It has also been shown \cite{fudenberg98a} that a stable steady state
of the replicator dynamics is a Nash equilibrium. A stable steady
state is one that, after suffering from a small perturbation, is
pushed back to the same steady state by the system's dynamics. These
states are necessarily Nash equilibria because if they were not then
there would exist some particular small perturbation which would take
the system away from the steady state. This correspondence was further
refined by Bomze \cite{bomze86a} who showed that an asymptotically
stable steady state corresponds to a Nash equilibrium that is
trembling-hand perfect and isolated. That is, the stable steady states
are a refinement on Nash equilibria---only a few Nash equilibria
can qualify. On the other hand, it is also possible that a replicator
dynamics system will never converge. In fact, there are many examples
of simple games with no asymptotically stable steady states.

While replicator dynamics reflect some of the most troublesome aspects
of learning in multiagent systems some differences are evident. These
differences are mainly due to the replication assumption. Agents are
not usually expected to replicate, instead they acquire the strategies
of others.  For example, in a real multiagent system all the agents
might choose to play the strategy that performed best in the last
round instead of choosing their next strategy in proportion to how
well it did last time. As such, we cannot directly apply the results
from replicator dynamics to multiagent systems. However, the
convergence of the systems' dynamics to a Nash equilibrium does
illustrate the importance of this solution concept as an attractor of
learning agent's dynamics.

\subsection{Evolutionary Stable Strategies}
\label{sec:evol-stable-strat}

An Evolutionary Stable Strategy (ESS) is an equilibrium concept
applied to dynamic systems such as the replicator dynamics system of
the previous section. An ESS is an equilibrium strategy that can
overcome the presence of a small number of invaders. That is, if the
equilibrium strategy profile is $\omega$ and small number $\epsilon$
of invaders start playing $\omega'$ then ESS states that the existing
population should get a higher payoff against the new mixture
($\epsilon\omega' + (1 - \epsilon)\omega$) than the invaders.

It has been shown \cite{taylor78a} that an ESS is an asymptotically
stable steady state of the replicator dynamics. However, the converse
need not be true---a stable state in the replicator dynamics does not
need to be an ESS. This means that ESS is a further refinement of the
solution concept provided by the replicator dynamics. ESS can be used
when we need a very stable equilibrium concept.

\section{Learning Agents}
\label{sec:learning-agents}

The theory of learning in games provides the designer of multiagent
systems with many useful tools for determining the possible
equilibrium points of a system. Unfortunately, most multiagent systems
with learning agents do not converge to an equilibrium. Designers use
learning agents because they do not know, at design time, the specific
circumstances that the agents will face at run time.  If a designer
knew the best strategy, that is, the Nash equilibrium strategy, for
his agent then he would simply implement this strategy and avoid the
complexities of implementing a learning algorithm.  Therefore, the
only times we will see a multiagent system with learning agents are
when the designer cannot predict that an equilibrium solution will
emerge.

The two main reasons for this inability to predict the equilibrium
solution of a system are the existence of unpredictable environmental
changes that affect the agents' payoffs and the fact that on many
systems an agent only has access to its own set of payoffs---it does
not know the payoffs of other agents. These two reasons make it
impossible for a designer to predict which equilibria, if any, the
system would converge to. However, the agents in the system are still
playing a game for which an equilibrium exists, even if the designer
cannot predict it at design-time. But, since the actual payoffs keep
changing it is often the case that the agents are constantly changing
their strategy in order to accommodate the new payoffs.

Learning agents in a multiagent system are faced with a \emph{moving
  target function problem} \cite{vidal:98a}. That is, as the agents
change their behavior in an effort to maximize their utility their
payoffs for those actions change, changing the expected utility of
their behavior. The system will likely have non-stationary
dynamics---always changing in order to match the new goal. While game
theory tells us where the equilibrium points are, given that the
payoffs stay fixed, multiagent systems often never get to those
points.  A system designer needs to know how changes in the design of
the system and learning algorithms will affect the time to
convergence.  This type of information can be determined by using CLRI
theory.

\subsection{CLRI Theory}
\label{sec:clri-theory}

The CLRI theory \cite{vidalclri} provides a formal method for
analyzing a system composed of learning agents and determining how an
agent's learning is expected to affect the learning of other agents in
the system.  It assumes a system where each agent has a decision
function that governs its behavior as well as a target function that
describes the agent's best possible behavior. The target function is
unknown to the agent. The goal of the agent's learning is to have its
decision function be an exact duplicate of its target function. Of
course, the target function keeps changing as a result of other
agents' learning. 

Formally, CLRI theory assumes that there are $N$ agents in the system.
The world has a set of discrete states $w \in W$ which are presented
to the agent with a probability dictated by the probability
distribution $\mathcal{D}(W)$. Each agent $i \in N$ has a set of
possible actions $A_i$ where $|A_i| \geq 2$. Time is discrete and
indexed by a variable $t$. At each time $t$ all agents are presented
with a new $w \in \mathcal{D}(W)$, take a simultaneous action, and
receive some payoff.  The scenario is similar to the one assumed by
fictitious play except for the addition of $w$.

\begin{figure}[tb]
  \centerline{
    \xymatrix{
      & &  \delta_i^{t+1}  \ar@{~}[rrrrd]^{e(\delta_i^{t+1})}
      & & & \\
      \delta_i^t \ar[rru]^{\txt{Learn}} \ar@{~}[rrrrr]_{e(\delta_i^t)} &&
      & & & \Delta_i^t \ar[r]_{\txt{Move}} & \Delta_i^{t+1}
      }}
  \caption{The moving target function problem.}
  \label{fig:learn-mas}
\end{figure}
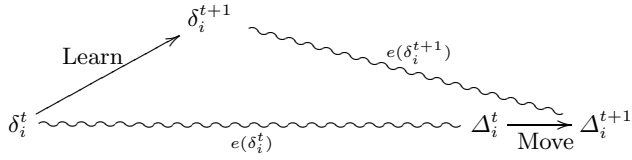

Each agent $i$'s behavior is defined by a decision function
$\delta_i^t(w) : W \rightarrow A$. When $i$ learns at time $t$ that it
is in state $w$ it will take action $\delta_i^t(w)$. At any time there
is an optimal function for $i$ given by its target function
$\Delta_i^t(w)$. Agent $i$'s learning algorithm will try to reduce the
discrepancy between $\delta_i$ and $\Delta_i$ by using the payoffs it
receives for each action as clues since it does not have direct access
to $\Delta_i$. The probability that an agent will take a wrong action
is given by its error $e(\delta_i^t) = \Pro[\delta_i^t(w) \neq
\Delta_i^t(w) \,|\, w \in \mathcal{D}(W)]$. As
other agents learn and change their decision function, $i$'s target
function will also change, leading to the moving target function
problem, as depicted in Figure~\ref{fig:learn-mas}.

An agent's error is based on a fixed probability distribution over
world states and a boolean matching between the decision and target
functions. These constraints are often too restrictive to properly
model many multiagent systems. However, even if the system being
modeled does not completely obey these two constraints, the use of the
CLRI theory in these cases still gives the designer valuable insight
into how changes in the design will affect the dynamics of the system.
This practice is akin to the use of Q-learning in non-Markovian
games---while Q-learning is only guaranteed to converge if the
environment is Markovian, it can still perform well on other domains.

The CLRI theory allows a designer to understand the expected dynamics
of the system, regardless of what learning algorithm is used, by
modeling the system using four parameters: Change rate, Learning rate,
Retention rate, and Impact (CLRI). A designer can determine values for
these parameters and then use the CLRI difference equation to
determine the expected behavior of the system.

The change rate ($c$) is the probability that an agent will change at
least one of its incorrect mappings in $\delta^t(w)$ for the new
$\delta^{t+1}(w)$. It captures the rate at which the agent's learning
algorithm tries to change its erroneous mappings. The learning rate
($l$) is the probability that the agent changes an incorrect mapping
to the correct one. That is, the probability that $\delta^{t+1}(w) =
\Delta^t(w)$, for all $w$. By definition, the learning rate must be
less than or equal to the change rate, i.e. $l \leq c$. The retention
rate ($r$) represents the probability that the agent will retain its
correct mapping. That is, the probability that $\delta^{t+1}(w) =
\delta^t(w)$ given that $\delta^t(w) = \Delta^t(w)$.

CLRI defines a volatility term ($v$) to be the probability that the
target function $\Delta$ changes from time $t$ to $t+1$. That is, the
probability that $\Delta^t(w) \neq \Delta^{t+1}(w)$. As one would
expect, volatility captures the amount of change that the agent must
deal with. It can also be viewed as the speed of the target function
in the moving target function problem, with the learning and retention
rates representing the speed of the decision function. Since the
volatility is a dynamic property of the system (usually it can only be
calculated by running the system) CLRI provides an impact ($I_{ij}$)
measure. $I_{ij}$ represents the impact that $i$'s learning has on
$j$'s target function. Specifically, it is the probability that
$\Delta_j^t(w)$ will change given that $\delta_i^{t+1}(w) \neq
\delta_i^t(w)$.

Someone trying to build a multiagent system with learning agents would
determine the appropriate values for $c$, $l$, $r$, and either $v$ or
$I$ and then use 
\begin{multline}
  \label{eq:main:simp}
    E[e(\delta_i^{t+1})] = 1 - r_i + v_i
    \left(
      \frac{|A_i|r_i - 1}{|A_i| -1}
    \right) \\
    + e(\delta_i^t)
    \left(
      r_i -l_i + v_i
      \left(
        \frac{|A_i|(l_i - r_i) + l_i - c_i}{|A_i| -1}
      \right)
    \right)
\end{multline}
in order to determine the
successive expected errors for a typical agent $i$. This equation
relies on a definition of volatility in terms of impact given by
\begin{equation}
  \label{eq:16}
  \begin{split}
       \forall_{w \in W} \; v_{i}^{t} &= \Pro[\Delta_i^{t+1}(w) \neq \Delta_i^t(w)] \\
       &= 1 - \prod_{j \in N_{-i}}(1 - I_{ji} \Pro[\delta_j^{t+1}(w) \neq \delta_j^t(w)]), \\
  \end{split}
\end{equation}
which makes the simplifying assumption that changes in agents'
decision functions will not cancel each other out when calculating
their impact on other agents. The difference equation
\eqref{eq:main:simp} cannot, under most circumstances, be collapsed
into a function of $t$ so it must still be iterated over.  On the
other hand, a careful study of the function and the reasoning behind
the choice of the CLRI parameter leads to an intuitive understanding
of how changes in these parameters will be reflected in the function
and, therefore, the system. A knowledgeable designer can simply use
this added understanding to determine the expected behavior of his
system under various assumptions. An example of this approach is shown
in \cite{brooks02a}.

For example, it is easy to see that an agent's learning rate and the
system's volatility together help to determine how fast, if ever, the
agent will reach its target function. A large learning rate means that
an agent will change its decision function to almost match the target
function. Meanwhile, a low volatility means that the target function
will not move much, so it will be easy for the agent to match it. Of
course, this type of simple analysis ignores the common situation
where the agent's high learning rate is coupled with a high impact on
other agents' target function making their volatility much higher.
These agents might then have to increase their learning rate and
thereby increase the original agent's volatility. Equation
\eqref{eq:main:simp} is most helpful in these type of feedback
situations.

\subsection{N-Level Agents}
\label{sec:n-level-agents}

Another issue that arises when building learning agents is the choice
of a modeling level. A designer must decide whether his agent will
learn to correlate actions with rewards, or will try to learn to
predict the expected actions of others and use these predictions along
with knowledge of the problem domain to determine its actions, or will
try to learn how other agents build models of other agents, etc. These
choices are usually referred to as n-level modeling agents---an idea
first presented in the recursive modeling method
\cite{gmytrasiewicz95a} \cite{gmytrasiewicz01a}.

A 0-level agent is one that does not recognize the existence of other
agents in the world. It learns which action to take in each possible
state of the world because it receives a reward after its actions. The
state is usually defined as a static snapshot of the observable
aspects of the agent's environment. A 1-level agent recognizes that
there are other agents in the world whose actions affect its payoff.
It also has some knowledge that tells it the utility it will receive
given any set of joint actions. This knowledge usually takes the form
of a game matrix that only has utility values for the agent. The
1-level agent observes the other agents' actions and builds
probabilistic models of the other agents. It then uses these models to
predict their action probability distribution and uses these
distributions to determine its best possible action. A 2-level agent
believes that all other agents are 1-level agents. It, therefore,
builds models of their models of other agents based on the actions it
thinks they have seen others take. In essence, the 2-level agent
applies the 1-level algorithm to all other agents in an effort to
predict their action probability distribution and uses these
distributions to determine its best possible actions. A 3-level agent
believes that all other agents are 2-level, an so on.  Using these
guidelines we can determine that fictitious play
(Section~\ref{sec:fictitious-play}) uses 1-level agents while the
replicator dynamics (Section~\ref{sec:replicator-dynamics}) uses
0-level agents.

These categorizations help us to determine the relative computational
costs of each approach and the machine-learning algorithms that are
best suited for that learning problem. 0-level is usually the easiest
to implement since it only requires the learning of one function and
no additional knowledge. 1-level learning requires us to build a model
of every agent and can only be implemented if the agent has the
knowledge that tells it which action to take given the set of actions
that others have taken. This knowledge must be integrated into the
agents. However, recent studies in layered learning \cite{stone00a}
have shown how some knowledge could be learned in a ``training''
situation and then fixed into the agent so that other knowledge that
uses the first one can be learned, either at runtime or in another
training situation. In general, a change in the level that an agent
operates on implies a change on the learning problem and the knowledge
built into the agent.

Studies with n-level agents have shown \cite{vidal:98b} that an
n-level agent will always perform better in a society full of
(n-1)-level agents, and that the computational costs of increasing a
level grow exponentially.  Meanwhile, the utility gains to the agent
grow smaller as the agents in the system increase their level, within
an economic scenario. The reason is that an n-level agent is able to
exploit the non-equilibrium dynamics of a system composed of
(n-1)-level agents.  However, as the agents increase their level the
system reaches equilibrium faster so the advantages of strategic
thinking are reduced---it is best to play the equilibrium strategy and
not worry about what others might do. On the other hand, if all agents
stopped learning then it would be very easy for a new learning agent
to take advantage of them. As such, the research concludes that some
of the agents should do some learning some of the time in order to
preserve the robustness of the system, even if this learning does not
have any direct results.

\section{Conclusion}
\label{sec:conclusion}

We have seen how game theory and the theory of learning in games
provide us with various equilibrium solution concepts and often tell
us when some of them will be reached by simple learning models.  On
the other hand, we have argued that the reason learning is used in a
multiagent system is often because there is no known equilibrium or
the equilibrium point keeps changing due to outside forces. We have
also shown how the CLRI theory and n-level agents are attempts to
characterize and predict, to a limited degree, the dynamics of a
system given some basic learning parameters.

We conclude that the problems faced by a designer of a learning
multiagent systems cannot be solved solely with the tools of game
theory. Game theory tells us about possible equilibrium points.
However, learning agents are rarely at equilibrium, either because
they are not sophisticated enough, because they lack information, or
by design. There is a need to explore non-equilibirium systems and to
develop more predictive theories which, like CLRI, can tell us how
changing either the parameters on the agents' learning algorithms or
the rules of the game will affect the expected emergent behavior.


\begin{thebibliography}{10}

\bibitem{bomze86a}
Bomze, I.:
\newblock Noncoopertive two-person games in biology: A classification.
\newblock International Journal of Game Theory \textbf{15} (1986)  31--37

\bibitem{brooks02a}
Brooks, C.H., Durfee, E.H.:
\newblock Congregation formation in multiagent systems.
\newblock Journal of Autonomous Agents and Multi-agent Systems (2002) to
  appear.

\bibitem{fudenberg90a}
Fudenberg, D., Kreps, D.:
\newblock Lectures on learning and equilibrium in strategic-form games.
\newblock Technical report, {CORE} Lecture Series (1990)

\bibitem{fudenberg98a}
Fudenberg, D., Levine, D.K.:
\newblock The Theory of Learning in Games.
\newblock {MIT} Press (1998)

\bibitem{gmytrasiewicz95a}
Gmytrasiewicz, P.J., Durfee., E.H.:
\newblock A rigorous, operational formalization of recursive modeling.
\newblock In: Proceedings of the First International Conference on Multi-Agent
  Systems. (1995)  125--132

\bibitem{gmytrasiewicz01a}
Gmytrasiewicz, P.J., Durfee., E.H.:
\newblock Rational communication in multi-agent systems.
\newblock Autonomous Agents and Multi-Agent Systems Journal \textbf{4} (2001)
  233--272

\bibitem{mitchell97a}
Mitchell, T.M.:
\newblock Machine Learning.
\newblock McGraw Hill (1997)


\bibitem{stone00a}
Stone, P.:
\newblock Layered Learning in Multiagent Systems.
\newblock {MIT} Press (2000)

\bibitem{taylor78a}
Taylor, P., Jonker, L.:
\newblock Evolutionary stable strategies and game dynamics.
\newblock Mathematical Biosciences \textbf{16} (1978)  76--83

\bibitem{vidal:98a}
Vidal, J.M., Durfee, E.H.:
\newblock The moving target function problem in multi-agent learning.
\newblock In: Proceedings of the Third International Conference on Multi-Agent
  Systems. (1998)

\bibitem{vidal:98b}
Vidal, J.M., Durfee, E.H.:
\newblock Learning nested models in an information economy.
\newblock Journal of Experimental and Theoretical Artificial Intelligence
  \textbf{10} (1998)  291--308

\bibitem{vidalclri}
Vidal, J.M., Durfee, E.H.:
\newblock Predicting the expected behavior of agents that learn about agents:
  the {CLRI} framework.
\newblock Autonomous Agents and Multi-Agent Systems (2002)


\end{thebibliography}

\end{document}